\documentclass[prx,twocolumn,prl,showpacs,superscriptaddress]{revtex4-1}
\usepackage{graphicx}
\usepackage{color}

\begin{document}

\title{First observation of flux avalanches in a-MoSi superconducting thin films}

\author{F. Colauto}
\affiliation{Departamento de F\'{\i}sica, Universidade Federal de S\~{a}o Carlos, 13565-905 S\~{a}o Carlos, SP, Brazil}

\author{M. Motta}
\affiliation{Departamento de F\'{\i}sica, Universidade Federal de S\~{a}o Carlos, 13565-905 S\~{a}o Carlos, SP, Brazil}

\author{A. Palau*}
\affiliation{Department of Materials Science, University of Cambridge, Pembroke Street, Cambridge CB2 3QZ, UK}

\author{M. G. Blamire}
\affiliation{Department of Materials Science, University of Cambridge, Pembroke Street, Cambridge CB2 3QZ, UK}

\author{T. H. Johansen}
\affiliation{Department of Physics, University of Oslo, POB 1048, Blindern, 0316 Oslo, Norway}

\author{W. A. Ortiz}
\affiliation{Departamento de F\'{\i}sica, Universidade Federal de S\~{a}o Carlos, 13565-905 S\~{a}o Carlos, SP, Brazil}

\date{\today}

\begin{abstract}
We have observed the occurrence of dendritic flux avalanches in an amorphous film of Mo$_{84}$Si$_{16}$. These events are understood to have a thermomagnetic origin and involve the abrupt penetration of bursts of magnetic flux taking place within a limited window of temperatures and magnetic fields. While dc-magnetometry allows one to determine the threshold fields for the occurrence of the thermomagnetic instabilities, magneto-optical imaging reveals the spatial distribution of magnetic flux throughout the sample. Conducting appropriate experiments, typical for this goal, avalanches were confirmed to be a characteristic of this material, ruling out the otherwise admissible possibility of an experimental artifact or a feature related to defects in the film. After the present observation, a-MoSi can be included in the gallery of superconducting materials exhibiting flux avalanches when in the form of thin films, a characteristic that must be carefully taken into consideration when one plans to employ films of those materials in applications.
\end{abstract}

\maketitle


\section{Introduction}

The very first observation of sudden penetration of magnetic flux in superconductors was reported in vanadium specimens by Schawlow \cite{Schawlow56} in 1956. Abrupt penetrations of flux, so-called flux jumps or flux avalanches, have been also observed in superconducting thin films by using magneto-optical imaging (MOI). This technique has revealed the dendritic shape of such avalanches for several materials: YBCO \cite{Leiderer93,Bazil2014}, Nb \cite{Duran95}, MgB$_2$ \cite{Johansen01}, Pb \cite{Menghini05}, Nb$_3$Sn \cite{Rudnev03}, YNi$_2$B$_2$C \cite{Wimbush04}, NbN \cite{Rudnev05}, and more recently in a-MoGe \cite{Motta13}. 

Flux avalanches have been understood as due to thermomagnetic instabilities \cite{Mints81,Wipf91}. They occur at low temperatures when the magnetic diffusion is faster than the thermal diffusion and the sample has no time to redistribute the heat generated by the vortex movement. Then the temperature increases locally, reducing the critical current density and allowing further flux penetration. The time variation of the magnetic flux inside the superconductor induces an electrical field, generating more heat, leading to a positive feedback and, consequently, to a macroscopic avalanche. The theoretical background that describes the instabilities in films combines Maxwell equations with heat diffusion \cite{Denisov06}, providing results in good agreement with the experiments \cite{Yurchencko07,Denisov06a}. Numerical simulations have reproduced qualitatively and quantitatively avalanche features in type II superconducting films \cite{Vestgarden11,Vestgarden12}, shedding light on the dynamics of flux avalanche and even indicating that the temperature during the avalanche evolution can be higher than the critical one.

As a matter of fact, flux avalanches are deleterious to the practical performance of superconducting films, ruining their screening and transport properties. An effective way to suppress avalanches is by placing a metallic layer in close contact to the superconductor \cite{Baziljevich02}. Since flux penetration into the sample means temporal variation of the magnetic field, Faraday's law applied to the metallic cover exposed to this abruptly varying flux implies the appearance of eddy currents which, in turn, react against and inhibit the occurrence of the avalanche \cite{Colauto10}.

One of the most attractive technological applications of films in superconducting electronics are nowadays the superconducting nanowire single photon detectors (SNSPD) \cite{Goltsman01}. Recently, Korneeva \textit{et al.} \cite{Korneeva13} and Verma and collaborators \cite{Verma14} have described the amorphous alloys of MoGe and MoSi, respectively, as promising systems for developing such devices. At temperatures of the order of a few Kelvin, detection efficiencies approaching 20\% were achieved. These amorphous films have some interesting superconducting properties: their critical temperature ($T_c$) increases with amorphicity \cite{Lehmann81}; the Ginzburg-Landau parameter $\kappa$ is approximately one order of magnitude larger than in typical materials, for instance, Nb \cite{Kubo88,Motta14}; and the density of intrinsic pinning centers is low \cite{Kubo88}. 

In the present work we use MOI and dc magnetization (DCM) measurements in order to investigate spontaneous dendritic flux avalanches in a-MoSi film. Both techniques are complementary in order to map the temperature and field ranges and the features of the flux avalanches \cite{Colauto07,Colauto08}. Our results confirm that amorphous MoSi belong in the list of materials exhibiting instabilities of thermomagnetic origin.

\section{Material and methods}

The amorphous thin film of the molybdenum compound Mo$_{84}$Si$_{16}$ was grown via ultra-high vacuum dc-magnetron sputtering on a (100) SiO$_2$ substrate cooled down to liquid nitrogen temperature (77 K). The film is roughly rectangular with an area of 2.2 mm x 2.7 mm and thickness of 250 nm. The critical temperature is 7.2 K with a transition width as sharp as 0.2 K, as determined by an ac-susceptibility measurement with amplitude and frequency of 0.1 Oe and 100 Hz, respectively. One of the techniques used here to investigate the instability regime was the DCM, which measures the overall magnetic moment of the sample through experiments carried out in a Quantum Design magnetometer MPMS-5S. The second technique employed in this work was MOI, which is based on the occurrence of the magneto-optical Faraday effect. This effect occurs when the plane of a linearly polarized beam of light is rotated when it passes through a Faraday-active material under a magnetic field applied parallel to the incident beam \cite{Jooss2002}. Since the superconductors have not shown this effect significantly, a Bi-doped yttrium iron garnet film (Bi:YIG) with in-plane magnetization \cite{Helseth2001} was placed on top of the superconducting specimen and employed as a sensor. When the magnetic flux penetrates into the sample, the magnetic domains in the indicator flip locally. Then, the reflected light rotates and passes through an analyzer that is aligned in 90 degree to a polarizer, revealing contrast between penetrated and flux-free regions of the sample in real-time. Our experimental setup has a spatial resolution of 2.2 $\mu$m for an amplification of 50 times, and a time resolution of the order of 100 ms.

\section{Results and discussions}

\begin{figure}[t!]
\centering
\includegraphics[width=8.7cm]{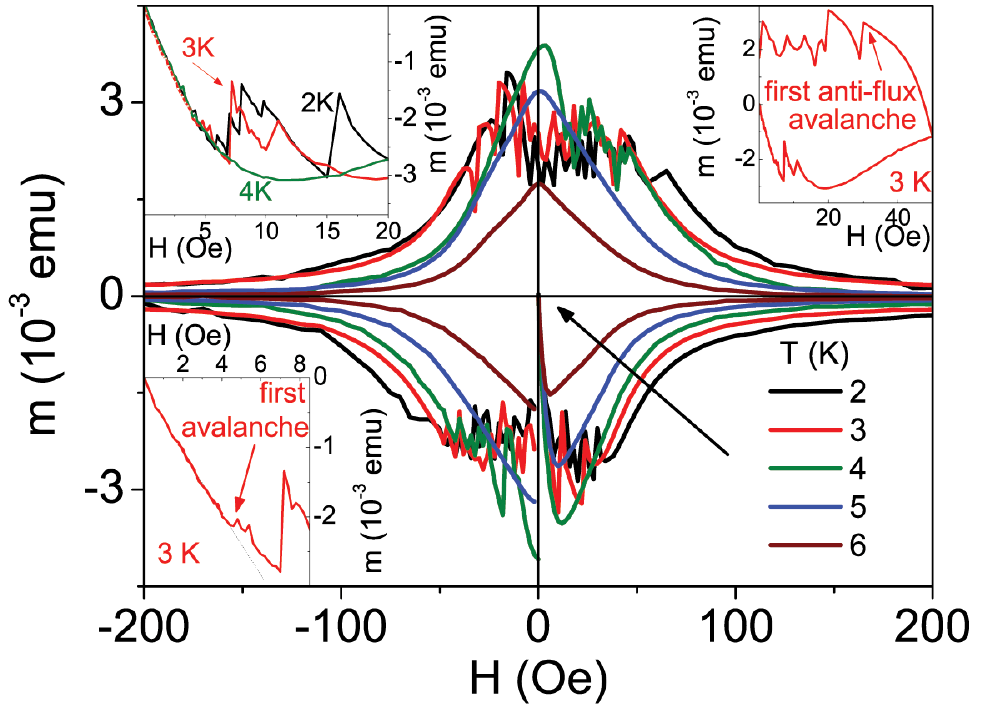}
\caption[width=8cm]{Isothermal magnetic hysteresis loops measured at several temperatures for an a-MoSi film. At lower temperatures, the strong noisy behavior around H = 0 Oe is the signature of flux avalanches. The top left-hand inset presents the virgin branches at 2 K and 3 K. At 4 K avalanches occur only when the external field is decreased. The bottom left-hand inset highlights the first avalanche in the virgin branch at 3 K that matches with the MO image shown ahead. The top right hand inset shows in detail the magnetization loop for positive fields taken up to 50 Oe at 3 K. The arrow in the main panel indicates the sense of increasing temperature.}\label{Figure1}
\end{figure}

Flux avalanches develop spontaneously when the magnetic field is applied perpendicular to superconducting films, for both increasing and decreasing the field amplitude. Fig. \ref{Figure1} shows magnetization loops for an a-MoSi film, where a noisy response is noticeable for lower temperatures. Both forward and backward branches exhibit flux jumps, including the virgin curves, as highlighted in the top left-hand inset. Similar to other materials, instabilities in a-MoSi take place in a certain range of fields and temperatures. Above a specific characteristic temperature, $T^*$, the magnetic response is entirely smooth as a function of the applied field, for increasing and decreasing fields. For the investigated sample, $T^*$ = 4 K, i.e., $T^*$ = 0.55$T_c$, whereas for Nb and MgB$_2$ films, this fraction is around 0.5$T_c$ and 0.25$T_c$, respectively \cite{Colauto07,Colauto08}.

In the virgin curves, the jumps begin around 5 Oe, as shown in the bottom left-hand inset for 3 K, and stop around 30 Oe, above which the magnetic moment is smooth again. The existence of such threshold fields can be derived from the thermomagnetic model \cite{Yurchencko07}. The lower limit is reached when a minimum flux penetration depth ($l^*$) occurs in the sample, then instabilities are likely to occur and trigger flux avalanches. Denisov \textit{et al.} \cite{Denisov06a} studied a set of strips with different widths and found out that $l^*$ = 82 $\mu$m for a MgB$_2$ film, which means if the half-width of the film is smaller than this value, no avalanches will occur. The instabilities cease when $l^*$, which depends on thermal and superconducting parameters, reaches the half-width of the specimen; this defines the upper threshold limit.

In addition, the fluctuations in the reversing branch restart around 50 Oe, slightly higher than the field above which the flux avalanches vanish in the virgin curves. At $T^*$, for instance, the noisy response takes place only for decreasing fields. A reasonable explanation relies on the fact that the slope of magnetic flux profile at the edges is larger while decreasing the applied field than that for increasing it. It means a higher critical current density and, consequently, the avalanches are triggered at higher fields \cite{Qviller10}.

\begin{figure}[t!]
\centering
\includegraphics[width=8.7cm]{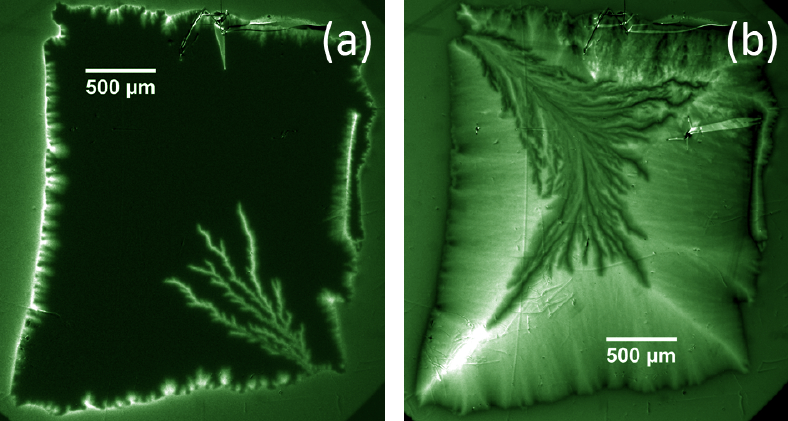}
\caption{MO images of an amorphous MoSi thin film taken at 3 K. Panel (a) presents the detection of the first avalanche at 4.5 Oe, after a ZFC procedure. Image (b) shows the first avalanche of antiflux, occurring at 20 Oe while the applied field is decreased after reaching a maximum of 50 Oe.}\label{Figure2}
\end{figure}

The magneto-optical (MO) images in Fig. \ref{Figure2} show how the avalanches develop into the sample. Panel (a) presents the detection of the first jump after a zero field cooling (ZFC) procedure, in a similar manner to that employed on DCM measurements. The morphology of the flux penetration is clearly dendritic, extending over about 10\% of the sample area. The applied field for the occurrence of this first avalanche taken by MOI at 3 K is 4.5 Oe, which is equivalent to that detected by DCM as can be observed in the bottom left-hand inset in Fig. \ref{Figure1}. Fig. \ref{Figure2}(b) was obtained in the same run and reveals the first jump at 20 Oe while decreasing the field after a maximum value of 50 Oe, which is also similar to the loop in the top right-hand inset in Fig. \ref{Figure1}. Moreover, the field interval within which avalanches occur is wider in the reverse branch than in the virgin curve. Regarding the magnitude of the magnetic moment, its fluctuations are related to the size of the avalanche, i.e., the higher is the variation of the magnetization, the larger is the dendrite grown into the sample \cite{Colauto08}. As a comparison with the above mentioned 10\% of the sample area occupied by the first dendrite in the virgin curve, the first anti-dendrite takes areal fraction of around 25\%. 

\begin{figure}[t!]
\centering
\includegraphics[width=8.7cm]{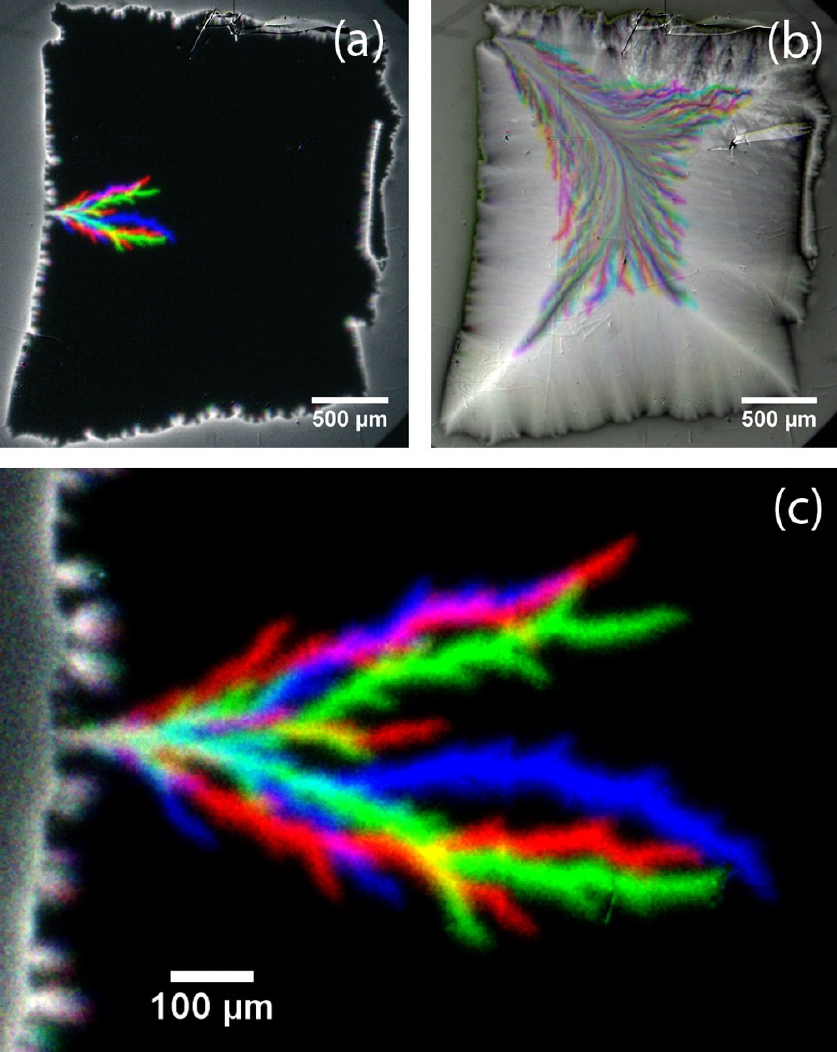}
\caption{Overlay of images taken from three independent MOI runs performed at identical conditions. The temperature measured is 2.65 K (system base temperature) reached after ZFC procedure. Each image received an RGB (red, green, blue) color in order to facilitate the observation. In the regions where flux penetration of the three measurements overlap the image is gray, otherwise it is colored. Panel (a) shows the first avalanches at 3 Oe and the colors indicate the avalanches develop through different ways into the sample. Panel (b) presents the first antiflux avalanches that enter the sample at 18 Oe, while the applied field is decreased after 50 Oe. Image (c) is a zoomup that shows details of the image shown in (a).}\label{Figure3}
\end{figure}

Flux avalanches do not cause any physical damage to the sample, which can be clean simply by driving it to the normal state. Edge defects are the most likely places where flux avalanches are triggered due to current crowding effects \cite{Clem11,Adami13} and strong local electric fields \cite{Vestgarden07}, however the abrupt phenomenon is stochastic as demonstrated by the images shown in Fig. \ref{Figure3}. Three experiments were performed within the same conditions, as follows: (i) the sample was cooled from a  temperature higher than $T_c$ down to the base temperature of the MOI setup (2.65 K) in a zero field environment; (ii) subsequently, the external magnetic field was increased up to 50 Oe in steps of 0.5 Oe and then decreased to zero in steps of 1.0 Oe, acquiring one image after each field variation. Each run received a different RGB (red, green, blue) color. Dark means flux-free regions, bright indicates the complete overlap of RGB colors and colored branches correspond to different paths traced by the penetrating flux. Fig. \ref{Figure3}(a) depicts the composition of the first avalanche at 3 Oe. An enlargement of the region where the dendrites take place can be observed in panel (c). The outcome image is colored, meaning that the avalanche is neither guided by internal defects nor related to any experimental artifact. Panel (b) shows the overlap of the first anti-dendrites at 18 Oe, while decreasing the field after a maximum field of 50 Oe. Around these anti-dendrites the overlapping of the images is colored, indicating irreproducibility in this region.

The occurrence of spontaneous flux avalanches seems to be a quite general feature of superconductors in the thin film geometry. The existence of field and temperature thresholds, and the dendritic nature of the avalanches in a-MoSi constitute signatures of the thermomagnetic origin of the events reported here. 

\section{Conclusions}
We have imaged dendritic flux avalanches in an amorphous superconducting thin film of MoSi. The avalanches develop spontaneously in form of dendrites, similar to other superconducting films, in a perpendicular applied magnetic field. The thermomagnetic avalanches arise for decreasing, as well as for increasing applied fields - including the virgin state - as long as the temperature is below $T^*$ = 4 K (0.55$T_c$). Flux avalanches are complex events which, once triggered, have unpredictable evolution and, as such, deserve special attention when superconducting films are considered in practical applications.

\section*{Acknowledgment}
This work was partially supported by the Brazilian funding agencies CAPES, CNPq and FAPESP (Grants 2007/08072-0 and 13/16097-3, Sao Paulo Research Foundation) and by the UK Engineering and Physical Sciences Research Council. WAO acknowledges support from CAPES (Grant AEX 3963/14-4). MM is grateful for the support from CNPq (Grant 151257/2014-0) and CAPES (AEX 4464/14-1).
\\

*Present address: Institut de Ciencia de Materials de Barcelona, E-08193 Bellaterra, Spain.


\end{document}